\documentclass[twocolumn,showpacs,preprintnumbers,amsmath,amssymb]{revtex4}

\usepackage{graphicx}
\usepackage{dcolumn}
\usepackage{bm}
\usepackage{amssymb}
\usepackage{enumerate}

\def\address{\@ifstar{\address@star}%
  {\@ifnextchar[{\address@optarg}{\address@noptarg}}}

\begin{document}

\author{S.N.~Gninenko$^{1,2,3}$}
\author{N.V.~Krasnikov$^{1,3}$}

\affiliation{$^1$ Institute for Nuclear Research of the Russian Academy of Sciences 117312 Moscow, Russia\\
$^{2}$ Millennium Institute for Subatomic Physics at  the High-Energy Frontier (SAPHIR) of ANID, \\ 
	Fern\'andez Concha 700, Santiago, Chile\\
$^{3}$ Joint Institute for Nuclear Research 141980 Dubna, Russia}



\title{Search for mirror hidden sector with neutral kaons at NA64}

\date{\today}

\begin{abstract}
Mirror matter from the dark hidden sector with the same particle content  and gauge interactions as in the standard model
 is still an interesting candidate for dark matter.  Several experiments on search for  positronium and neutron oscillations
 into their mirror partner have been conducted recently.  
In this work we  consider the  transitions $K^0-K^0_m$ of a neutral kaon into a hidden mirror kaon. It is shown that their probability 
can be probed with the sensitivity of $P(K^0-K^0_m )\lesssim 10^{-7}$ from the search for missing-energy  events 
in the NA64  experiment at the CERN SPS with  $\sim 10^{11}$ $K^+$ on target. This would result in a limit on the $K^0 - K^0_m$ mixing parameter  much stronger than the similar bound estimated from the cosmological considerations.

\end{abstract}
\maketitle

\section{Introduction}
Mirror matter from the dark mirror sector with a particle content and gauge interactions, which is a mirror copy of the standard model, is still a valuable candidate for dark matter, for a recent review, see \cite{mohentr}, and also \cite{zbmm1,  zbmm2, zbmm3}.
In this class of models the mirror symmetry could be either unbroken, resulting in the mirror  particles  having the same masses  as in the Standard Model (SM) 
\cite{ly,kobz,blin,volkas,zb1,foot1,okun}, or  (ii) The symmetry could be  spontaneously broken and mirror sector particles have  masses which are similar compared to the SM particles\cite{zbmoh,zbdolg}.
Already long ago,  Kobzarev, Okun and Pomeranchuk \cite{kobz} found out that the interaction between the SM and mirror particles cannot be  strong or electromagnetic.  They 
 interact presumably  via gravity, and possibly by other feeble forces. Such new interactions, induced e.g. by
the photon-mirror photon kinetic mixing \cite{okunmix, holdom} would result in  positronium - mirror positronium oscillations as was first pointed out by Glashow \cite{glashow}, see also \cite{sng1,footsng}.  Several experiments  motivated by this observation searched for the effect  \cite{atoyan, mitsu, bader, paolo, paolo1, paolo2},  see also \cite{zblepidi}.
Another possibility proposed in Ref.\cite{zbbento} is the neutron mixing with the mirror neutron resulting in 
 oscillations between the two states. This process has been looked for in experiments \cite{psinn, ser, alt, ill1, psi1, psi2, ill2} that provided a stringent  upper limit on the mixing parameter for the case of unbroken mirror symmetry, see, e.g., Refs.\cite{ill1, psi1}, and also study the case when neutron and mirror neutron have substantial mass splitting, i.e. the case of broken mirror symmetry, see, e.g.   recent result from the experiment at the ILL reactor \cite{ill2}.
 Mirror model for sterile neutrinos giving a subdominant contribution to solar neutrino physics was studied  in Ref.\cite{fv, berez}.

The possibility of $K^0$ to  mirror $K^0_m$ transition was first discussed by  Nikolaev and Okun  \cite{no}. They proposed to 
 search for these transformations in an experiment a'la "light shining through the wall". Namely, the primary $K^0$ beam producing a flux of mirror kaons 
 due to the $K^0 - K^0_m$ conversion in flight  is fully absorbed in a shield installed upstream of a detector. The signal in the detector is produced by the decays of  
 the ordinary  $K^0$'s regenerated in the beam 
 due to  the $K^0_m - K^0$ transitions of mirror kaons  that are sterile and penetrated the shield without interactions. In this approach, the signal rate is proportional to $\sim a^4$, 
 here $a$ is  the amplitude of the  $K^0\to K^0_m$ conversion,  where one small  factor $a^2$ came from the   $K^0 \to K^0_m$ transition before the shield, and another small factor 
  $a^2$ is  from  the  $K^0_m \to K^0$ regeneration  after the shiled.
 Below,  we  consider another  type of the experiment based on the active target plus missing-energy approach developed for the search for  invisible decays of dark photons
\cite{sngldms, yua}. Its advantage is that the signal rate is $\sim a^2$, thus much less kaons on target are required to reach the same level of 
sensitivity as in the proposal of Ref.\cite{no}. 
\par The rest of the paper is organized as follows. The phenomenology of the  $K^0-K^0_m$ system is  described  in Sec. II.
The experimental method, the  setup,   and   the expected sensitivity  of the proposed search  are  presented in Sec. III. 
Cosmological constraints are  discussed in Sec. IV.
Sec.V summarises and provides concluding remarks. 

\section{Phenomenology of the $K^0-K^0_m$ system}

Let us introduce a mirror world of particles by assuming that the total Lagrangian is invariant under 
$CPM$ transformation. Here, $C$, $P$, $M$ are charge conjugation, space inversion, and an  $ordinary~particles \leftrightarrow mirror~particles$ transformation. 
The following states with definite $CP$ parities  for the ordinary  and mirror world can be introduced:
\begin{eqnarray}
K_1= \frac{K^0+\overline{K}^0}{\sqrt{2}},~K_{1m}= \frac{K^0_m+\overline{K}^0_m}{\sqrt{2}}, ~ CP =+1 \\ 
K_2= \frac{K^0-\overline{K}^0}{\sqrt{2}}; K_{2m}= \frac{K^0_m-\overline{K}^0_m}{\sqrt{2}},~ CP =-1,  
\end{eqnarray}
and  the following states with definite $CPM$ parities  
\begin{eqnarray}
K^{+}_{1}= \frac{K_1+ K^0_{1m}}{\sqrt{2}},~K^{+}_{2}= \frac{K_2 - K_{2m}}{\sqrt{2}},~ CPM =+1 \\ 
K^{-}_{1}= \frac{K_1- K^0_{1m}}{\sqrt{2}},~K^{-}_{2}= \frac{K_2 + K_{2m}}{\sqrt{2}},~ CPM =-1 
\end{eqnarray}
which are associated with  particles of definite and distinct mass $m^+_1$,  $m^+_2$, $m^-_1$, and $m^-_2$, and 
mean lifetimes $\tau^+_1$,  $\tau^+_2$, $\tau^-_1$, and $\tau^-_2$.

\par We know that in the ordinary world   only $CPT$ appears to be an exact symmetry of nature, while $C,~P$ and $T$ are known to be violated.
Within the Wigner-Weisskopf approximation, the time evolution of the neutral kaon system is described by:
\begin{equation}
i\frac{d\Phi(t)}{dt}=H\Phi(t)=\Bigl(M-i\frac{1}{2}\Gamma\Bigr) \Phi(t)
\label{kk}
\end{equation}
where $M$ and $\Gamma$ are $2\times 2$  Hermitian matrices, which are time independent,  and $\Phi(t)$ is a two-component state vector in the $K^0 - \overline{K}^0$ space. Denoting by $m_{ij}$ and $\Gamma_{ij}$ the elements of $M$ and $\Gamma$ in the  $K^0 - \overline{K}^0$basis, $CPT$ invariance implies
\begin{eqnarray}
m_{11}=m_{22}~~ (\rm{or} ~m_{K^0}=m_{\overline{K}^0})~ \rm{and} \\ \nonumber
 \Gamma_{11}=\Gamma_{22}~~  (\rm{or} ~\Gamma_{K^0}=\Gamma_{\overline{K}^0})
\end{eqnarray}
The eigenstates of Eq. (\ref{kk}), called  $K_{S}$ and $K_{L}$, are given by 
\begin{equation}
K_S= \frac{1}{\sqrt{2(1+|\epsilon_{S}|^2)}}\Bigl(K_1 + \epsilon_S K_2\Bigr) 
\label{eq:kses}
\end{equation}
\begin{equation}
K_L= \frac{1}{\sqrt{2(1+|\epsilon_{L}|^2)}}\Bigl(K_2 + \epsilon_L K_1\Bigr) 
\label{eq:kles}
\end{equation}
where, assuming $CPT$ conservation, the parameter $\epsilon=\epsilon_S=\epsilon_L$ describing  $CP$ violation, in 
terms of the elements of $M$ and $\Gamma$, is 
\begin{equation}
\epsilon=\frac{{\rm Im}(\Gamma_{12})+i{\rm Im}(M_{12})}{\Gamma_S-\Gamma_L-2i\Delta m}
\end{equation}
where $\Gamma_S(\Gamma_L)$ is the $K_S(K_L)$ decay rate, and $\Delta m$ is the $K_L - K_S$ mass difference. 

\par Assuming existence of the mirror matter,  similar to Eqs.(\ref{eq:kses}, \ref{eq:kles}), the short-lived and long-lived states 
can be defined as \cite{no}:
\begin{equation}
K^{+(-)}_{S}= \frac{1}{\sqrt{2(1+|\epsilon^{+(-)}|^2)}}\Bigl(K^{+(-)}_{1} + \epsilon^{+(-)} K^{+(-)}_{2}\Bigr)  
\label{eq:ks+-es}
\end{equation}
\begin{equation}
K^{+(-)}_{L}= \frac{1}{\sqrt{2(1+|\epsilon^{+(-)}|^2)}}\Bigl(K^{+(-)}_{2} + \epsilon^{+(-)} K^{+(-)}_{1}\Bigr) 
\label{eq:kl+-es}
\end{equation}
with the corresponding eigenvalues $\lambda^{+(-)}_{S,L}= m^{+(-)}_{S,L} - i\frac{1}{2}\Gamma^{+(-)}_{S,L}$. Note, that $\lambda^+_{S,L} \neq \lambda^-_{S,L}$.

\subsection{$K^0 - K^0_m$ mixing}
 Let us consider  the phenomena associated with $K^0 - K^0_m$ mixing.  
   Suppose we produce  $K^0$ in an experiment at $t = 0 $ in a typical strong charge-exchange  reaction:
\begin{equation}
K^+ +n \to K^0 +p 
\label{eq:react}
\end{equation}
by irradiating a target followed by a hermetic detector that absorb all secondary particles in the final state and located  at some distance from it.
The energy deposit in the detector by $K^0$ or its decay product is  equal to the primary $K^+$ energy as the   recoil proton energy  is  typically much  smaller. 
\par Taking into account Eqs.(2-12) the  $K^0$ state at $t= 0$ is then
\begin{equation}
\begin{split}
&|\psi(0)>=|K^0>=\frac{1}{\sqrt{2}}[ (1-\epsilon^+)(|K^+_S>+|K^+_L>) \\
&+ (1-\epsilon^-)(|K^-_S>+|K^-_L>)] 
\end{split}
\end{equation}
For simplicity, let us  ignore $CP$ violation
\begin{equation}
\begin{split}
&|\psi(0)>=|K^0>=\frac{1}{\sqrt{2}}[ (|K^+_S>+|K^+_L>) \\
&+ (|K^-_S>+|K^-_L>)] 
\end{split}
\end{equation}
After a propagation time $t$  along the kaon beam to the detector, the wave function of $K^0$ is 
\begin{equation}
\begin{split}
&|\psi(t)>=\frac{1}{\sqrt{2}}[ (|K^+_S(t)>+|K^+_L(t)>) \\
&+ (|K^-_S(t)>+|K^-_L(t)>)] = \\
&\frac{1}{\sqrt{2}}[ (|K^+_S(t)>e^{-\lambda^+_St}+|K^+_L(t)>e^{-\lambda^+_Lt}) \\
&+ (|K^-_S(t)>e^{-\lambda^-_St}+|K^-_L(t)>e^{-\lambda^-_Lt})] 
\end{split}
\end{equation}
If the $K^0$ time of flight  is much greater that the lifetime of the short-lived components, 
$t \gg \tau^{+,-}_S=1/\Gamma^{+,-}_S$, we have 
\begin{equation} 
|\psi(t)>=\frac{1}{\sqrt{2}}e^{-i\lambda_L t}(|K_L(t)>+|K_{Lm}(t)>)
\end{equation}
where 
\begin{equation} 
\begin{split}
&|K_L(t)> \approx |K_2(t)>, |K_{Lm}(t)>\approx i\delta \lambda_L t  |K_{2m}(t)>\\
&\lambda_L = \frac{1}{2}(\lambda^+_L + \lambda^-_L), \delta \lambda_L = \frac{1}{2}(\lambda^+_L - \lambda^-_L)
\end{split}
\end{equation}
The mirror $K^0_m$  fraction of the beam  for $t \ll \tau^{+,-}_L=1/\Gamma^{+,-}_L$, is just
\begin{equation}
\begin{split}
&f(K_{2m}) = |<K_{2m}| |\psi(t)>|^2  \\
&\frac{1}{4}e^{-(\Gamma^+_L+\Gamma^-_L)t/2}|\delta \lambda_L|^2 t^2 \approx \frac{1}{4}|\delta \lambda_L|^2 t^2 
\end{split}
\label{eq:frac}
\end{equation} 
The mirror component is sterile and, hence, would penetrate the detector without interaction resulting in a striking 
experimental signature - complete disappearance of the primary beam energy in the experiment with the signal sensitivity $\propto |\delta \lambda_L|^2 t^2$. As mentioned in Sec. I, 
this  makes the approach  based on searching for events with a large missing energy, see Sec. III, 
 more advantageous compared to the one  based on the  "light-shining-through the wall" technique proposed in Ref.\cite{no}.
The sensitivity of the latter is $\propto |\delta \lambda_L|^4$ due to the additional small  extra factor $\propto |\delta \lambda_L|^2$ related to the regeneration of the ordinary $K^0$'s. 
\subsection{Dumping of the $K^0_m$ wave function} 
Let us assume that in  an  experiment  $K^0$  is  formed in a target
of a material and than propagetes  in air not in vacuum. After ejection from the target, the $K^0$ meson can lose its energy due to elastic collisions, while 
the mirror component $K^0_m$ is sterile and does not interact with the air.  The significant difference in properties of  $K^0$ and 
 $K^0_m$ could  leads to the effect that after each collision the $K^0-K^0_m$ system became an incoherent mixture of $K^0$  and $K^0_m$.
It  is easy to show that if one starts with pure $K^0$ state, the probability of $K^0-K^0_m$ conversion 
 will be 
\begin{equation}
P(K^0 \to K^0_m) \simeq \frac{|\Delta \lambda_L|^2 t^2}{2(\omega_c t) }
\end{equation}
where $\omega_c$ is the $K^0$  collision rate in air. Thus, the $K^0 \to K^0_m$ transition rate will be 
reduced by a factor $N_{col} \simeq \omega_c t$ compared to the vacuum rate, where $N$ 
is the number of $K^0$-nuclei collisions during the propagation time to the detector $t$.
\par Let us estimate the average number of $K^0$ collisions in the air. 
The average namber of collisions is determined by the formula
\begin{equation}
  N_{col} = N_{av} \rho_{air} \sigma_{tot} L
  \label{Ncol}
\end{equation}
Here $N_{av} = 6.02  \cdot 10^{23} ~mole^{-1}$ is the Avogadro number, $\rho_{air}$ is the density of air,
$\sigma_{tot}$ is the cross section of the $K^0$ meson on the ``air moleculae''
(mainly the mixing of  $O_2$ and $N_2$ moleculae) and the $L$ is the length of the
$K^0$ meson propagation. In our estimates we take $L = 5~m$, $\sigma_{tot} = 20 ~mb $
and the air density at room temperature $\rho_{air} =  1.3 \cdot 10^{-3} ~g/cm^3$.
We take  the average atomic number of the ``air moleculae''
$A_{air} = 29$. For estimation of Kaon ``air moleculae'' cross section we use the  the additive approximation
$\sigma_{tot}(K^0~air) =A_{air} \sigma_{tot}(K^0 ~nucleon )$. 
As a consequence of the formula (\ref{Ncol}) we find the average number of collisions is
\begin{equation}
  N_{col} \approx 8 \cdot 10^{-3}
\label{Ncolbound}
\end{equation}
It means that the effects of $K^0$ collisions are small and we can neglect them in the study of
the  $K^0-K^0_m$ transitions.   

\begin{figure*}[tbh!]
\includegraphics[width=.9\textwidth]{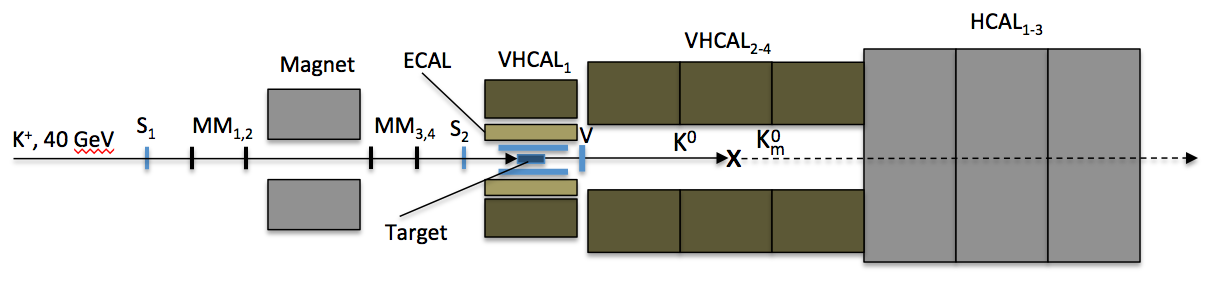}
\caption{\label{fig:setup} 
Schematic illustration of the NA64 setup to search for the $K^0 \to K^0_m$ conversion in flight (see text).   }
\label{setup}
\end{figure*}   

\section{An experiment to search for  $K^0-K^0_m$ transformation} \label{sec:ExpInvisible}
The  detector specifically designed to search for the $K^0-K^0_m$ transitions is schematically shown in 
Fig. \ref{setup}. This experimental setup  is complementary to the one proposed for the search for
invisible decays of dark photons at the SPS at CERN \cite{sngldms,yua}.    
The experiment could employ, e.g., the  H4 hadron  beam, which is produced in the target T2 of the CERN SPS and transported to the detector in an evacuated beamline tuned to a freely adjustable  beam momentum from 40-100 GeV/c \cite{sps}. 
The typical maximal beam  intensity at $\simeq$ 50-100 GeV, is of the order of $\simeq 10^7~\pi^\pm$ 
and $\simeq 3\times 10^5~ K^\pm$ for one  SPS spill with $10^{12}$ protons on target. The typical SPS cycle for 
fixed-target (FT) operation lasts 14.8 s, including 4.8 s  spill duration. The maximal number of FT cycles is four per minute.  
\par The beam of incident charged kaons is defined by the scintillating counters S$_{1,2}$. The momentum of the beam is additionally selected  with a momentum spectrometer
consisting of a dipole magnet and a low-density tracker, made of a set of  Micromegas chambers MM$_{1-4}$.   
The $K^0$s are produced in beam kaons scattering off nuclei 
in the active target (T). The T is  surrounded by the veto system consisting of of a set of high-efficiency scintillating counters (V),  an
 electromagnetic calorimeter (ECAL) and a veto hadronic calorimeter (VHCAL$_1$)   used against photons or charged secondaries that could escape the target at a large angle. 
The three VHCAL$_{2-4}$ modules are used to form a decay volume DV of the length $L=5$ m for development  of the $K^0->K^0_m$ transformations. In addition, VHCAL$_{2-4}$ modules and three massive hadronic calorimeter module (HCAL$_{1-3}$) located at the end of the setup are used as a fully hermetic calorimeter system to detect energy deposited by secondaries from the  primary  interactions $K^+ A \to$ anything  of $K^+$s with nuclei $A$  in the target. 
For searches at low  energies,  Cherenkov counters  to enhance the incoming hadron tagging efficiency can be used. 

\par The method of the search is as follows. The source of $K^0$s is the charge-exchange reaction  \eqref{eq:react} of high-energy kaons on nucleons of an  active target
where the neutral kaon is emitted mainly in the forward direction with the beam momentum and the 
recoil nucleon carries away a small fraction of the beam energy.
 If the process of $K^0\to K^0_m$ transitions exist, the $K^0_m$ component produced in the beam during time of flight $t(=L/c\gamma~{\rm in}$$~K^0~{\rm c.m.s.})$  would   penetrate the rest of the detector without   
interaction and any energy deposition in the HCAL. 
The occurrence of $K^0\to K^0_m$ transition produced in $K^\pm$ interactions would appear as an excess of events with a signal in the $T$, see Fig.~\ref{setup} and zero energy deposition in the rest of the detector (i.e. above that expected from the background sources). 
\subsection{Expected sensitivity}
To estimate the sensitivity of the proposed experiment 
 a simplified feasibility study  based on GEANT4 \cite{geant}
Monte Carlo simulations have been  performed for 40 GeV kaons. In this simulations the ECAL is an array of $3\times 3$ lead-scintillator  counters of the Shashlyk type  ($X_0 \simeq 2$ cm) (see, e.g. Ref.\cite{yua}),  each with the size of $36\times 36 \times 400$ mm$^3$, allowing for accurate measurements of the lateral energy leak from the target. The target itself is the ECAL central cell with  thickness $\simeq 0.1\lambda_{int}$, the surrounding counters serves as the veto counters which should detect a small energy deposit.
A new high-resolution active target made of lead tungstate (PbWO4) crystals \cite{pwo} has been also considered. 
The veto system's inefficiency  for the MIP (minimum ionising particle)  detection  is conservatively  assumed to be $\lesssim 10^{-4}$. 
Each VHCAL module is a  sandwich of 40 alternating layers of copper and scintillator with thicknesses of 25 mm and 4 mm,  
respectively, and with a lateral size of $60\times 60$ cm$^2$ and the central hole of $12\times 12$ cm$^2$.  
Each HCAL module consists of 48 alternating layers of iron and scintillator with thicknesses of 25 mm and 4 mm,  
respectively, the lateral size of $60\times 60$ cm$^2$ and 
a total thickness of $\simeq 7\lambda_{int}$.  The hadronic energy resolution of the HCAL calorimeters as a function of the beam energy is taken to be 
$\frac{\sigma}{E} \simeq \frac{ 60 \%}{\sqrt{E}}$. The energy threshold for the zero-energy in the HCAL is 0.1 GeV. 
\par To estimate the expected sensitivities simulations are used  to calculate fluxes and energy distributions  of mesons produced in the target by taking into account  
the relative normalization of the yield of  $K^0$ from the Refs. \cite{yud1, yud2}. The $K^0$  production cross section of the reaction \eqref{eq:react}
$\frac{d \sigma(K^+n\to K^0 p)}{dq} \simeq \frac{\sigma(K^-p\to \overline{K}^0 n)}{dq}$, and the  latter  can be expressed as \cite{yud1}
\begin{equation}
\begin{split}
&\frac{\sigma(K^-p\to \overline{K}^0 n)}{dq} = (1-Gt)(exp[c_\rho q]+R^2exp[c_A q ]) \\
&-2R[cos\phi_+ -GT cos\phi_-]exp[(c_\rho + c_A)q/2], 
\end{split}
\label{crsec}
\end{equation}
where  $q$ is the four-momentum transfer squared, $G= (33.5\pm 1.3)$ GeV$^{-2}$, $c_\rho=(15.5\pm0.3)$ GeV$^{-2}$, $c_A=(8.8\pm0.1)$ GeV$^{-2}$, $R=0.83\pm 0.05$, 
$cos\phi_+ = -0.08\pm 0.07$, and $cos\phi_+ = 0.23\pm 0.02$.
This  parametric form gives the charge-exchange cross sections  
for the $K^0$ production over the full phase space, up to $|q| \gtrsim 0.3$ GeV$^{-2}$. The  $K^0$  production cross sections in the target  were calculated  
from the linear extrapolation of \eqref{crsec} to the target atomic number. For the total number of incident kaons $n_{K^+}\simeq 10^{11}$ with the energy $E_0 = 40$ GeV, the number of produced $K^0$  is  $n_{K^0} \gtrsim 10^{7}$ which are emitted in the forward direction with the energy $E_{K^0} \simeq E_0$.  
The calculated fluxes and energy distributions  of mesons produced in the target are used to predict the number of signal events in the detector. 
\par For a given number of primary kaons $n_{K^+}$, the expected total number of mirror 
$K_{2m}$  kaons produced  within the decay volume of a  length $L$ and resulting in zero energy deposition in the detector  is given by  
\begin{eqnarray}
&n_{K_{2m}}= k n_{K^+}\cdot \int\frac{\sigma(K^+n\to K^0 p)}{dq} \times \nonumber \\
& \Bigl[1-{\rm exp}\Bigl(-\frac{L}{c\tau_{K_{L}}\gamma}\Bigr)\Bigr] P(K^0\to K_{2m}) \zeta \epsilon_{tag}  dq\nonumber \\
& \simeq  \zeta \epsilon_{tag} P(K^0\to K_{2m}) n_{K^0}
\label{nev}
\end{eqnarray}
where coefficient $k$ is a normalization factor that was
tuned to obtain the total  cross section of the meson production,  $P(K^0\to K_{2m}) =\frac{1}{4}|\delta \lambda_L|^2 t^2  $,  see Eq.(\ref{eq:frac}), 
and, neglecting the $K_S$ component,  $c\tau_{K_L}\gamma$ is the  $K_L$ decay length, respectively, $\zeta \simeq 0.9$ is the signal reconstruction efficiency, $\epsilon_{tag}\gtrsim 90\%$ is the tagging efficiency of the final state,  and $n_{K^0}$ is the total number of the produced $K^0$s. 
The estimation of background based on the analysis performed in Ref. \cite{Gninenko} and taking into account  the recent results of NA64 on the search for the 
$\eta, \eta' \to invisible$ decays  with the similar signature \cite{na64h},  results in  the  sensitivity for  signal events with the full missing energy to be at the level 
$\lesssim 10^{-11}$ per incoming kaon. Thus, in the case of no signal observation  the expected sensitivity of the proposed search for $K^0\to K^0_m$ conversion.
can be defined by using the relation $n_{K_{2m}} < n^{inv}_{90\%} $, where $n^{inv}_{90\%}$ (= 2.3 events) is the 90$\%$ C.L. upper limit for the  number of signal events.  Using Eq. (\ref{nev})  the limit for the conversion probability is 
\begin{equation}
P(K^0 \to K^0_m) \lesssim 10^{-7}
\label{eq:prob}
\end{equation}
for  $\simeq 10^{11}$ incident kaons.
This results in a bound on the $K^0 - K^0_m$ mixing parameter 
 \begin{equation}
 |\delta \lambda_L| \lesssim 3\times 10^{-19}~GeV, 
 \label{eq:lim}
 \end{equation}
which  is four orders of magnitude less compared to the value of the  $K_L - K_S$ mass difference \cite{pdg}. 
The statistical limit on the sensitivity of the proposed experiment is mostly set  by the number of 
accumulated events.   It could be further  improved by utilising a  more careful design of the experiment. For example, the primary reaction $\pi^- + p \to K^0 + \Lambda^0$ 
which has larger cross section,  can be used as a source of well-tagged $K^0$'s allowing to improve limits  of  \eqref{eq:prob} and \eqref{eq:lim} by an oder of magnitude.
 In the case of the signal observation,  several methods could be used to cross-check the result. For instance, 
 one could perform  measurements with different DV lengths.  In this case the expected number of signal events should vary as $n_S \sim t^2$.

\section{Cosmological constraints}
 Consider now cosmological bounds on  the mixing parameter $\delta \lambda_L$ \cite{referee}. Note, that  similar estimates have been performed in Ref.\cite{zbbento} for
the case of neutron- mirror neutron oscillations, see also \cite{zblepidi}.
   For the effective four fermion interaction
 \begin{equation}
   L_{eff} = \frac{1}{M_{X}^2}[ \bar{d}\gamma^{\mu}\gamma_5 s\bar{d}^m\gamma_{\mu}\gamma_5 s^m + h.c.]
   \label{efeq}
 \end{equation}
 connecting our $K_L$ mesons and mirror $K^m_L$ mesons the equlibrium temperature between our and mirror worlds
 is determined by the equation
 \begin{equation}
   n\sigma = H \,,
   \label{eqequil}
 \end{equation}
 where $ H = k_{H} \frac{T^2_{eq}}{M_{PL}} $
 (  $ k_H = 1.7 g^{1/2}_{*} $, $ g_{*} \sim 100 $) is the Hubble expansion rate, and $n = k_n T^3_{eq} ~(k_n = O(1)) $,
 $\sigma = k_{\sigma} \frac{T^2_{eq}}{M^4_{X}} ~(k_{\sigma} = O(1)) $ are the density and annihilation cross section.
 As a consequence of the equation (\ref{eqequil}) we find
 \begin{equation}
   T_{eq} = (\frac{M^4_Xk_H}{M_{PL}k_nk_{\sigma}})^{1/3} \,.
   \label{tempeq}
 \end{equation}
 From the requirement that the equilibrium temperature  $t_{eq}$ is larger the reheating temperature $T_r = 0.5$ TeV, 
 $T_{eq} \geq T_r$  \cite{zbbento}, we find that
\begin{equation}
  M_X \geq \frac{O(1)}{3} 10^{7}~{\text GeV} \,.
  \label{MXlimit}
  \end{equation}
  For the effective interaction (\ref{efeq}) the mixing parameter
  (\ref{eq:lim}) is  $\delta\lambda_L \sim \frac{f^2_Km_K}{M^2_X}$.  Here, $f_K \simeq0.16$ GeV is the kaon decay constant. 
    As a consequence of the inequality (\ref{MXlimit}) we find
 \begin{equation}
  |\delta \lambda_L| \leq O(1) 10^{-15} ~{\text GeV}
  \label{Klimit}
  \end{equation}
   The obtained bound \eqref{Klimit} is much weaker than the
  estimated projection  bound  (\ref{eq:lim}) from the NA64 experiment.
  
\section {Summary}
Mirror matter from the dark hidden sector with identical particle  and gauge interactions content as the standard model
 is still an interesting candidate for dark matter. 
Due to their specific properties, neutral kaons  are 
 still one of the most interesting probes of physics beyond the standard model  from both 
 theoretical and  experimental viewpoints. 
 In this work, the transformations of  an ordinary neutral kaon $K^0$-mesons into a sterile mirror $K^0_m$ meson  are  considered. We 
 show that their probability can be probed with the sensitivity  $P(K^0-K^0_m )\lesssim 10^{-7}$  for  $\simeq 10^{11}$ $K^+$ mesons on target
  in the NA64  experiment at the CERN SPS using the missing-energy  signature. Interestingly, the estimated projection  bound  on the mixing parameter (\ref{eq:lim}) from the NA64 experiment is much stronger than the similar bound \eqref{Klimit} estimated from the cosmological considerations. 
 The transition probability is proportional to the square of the $K^0$ time of flight, $\sim t^2$, 
allowing an effective cross-check of the result in the case of a signal observation.

\par  As the experimental signature of the signal is identical to the one from  $K^0\to invisible$ transitions, the proposed experiment can  also search for  
$K_{S,L} \to invisible$ decays, which  are a powefrul tool for probing  new physics in the kaon sector \cite{gk1, gk2, gk3}. Note, that these decays modes have  never been sensitively  tested. Only recently, the  BESIII collaboration had set  first modest bounds on their existence \cite{bes}. 
 It is also important to stress, that the  first results from a proof-of-concept search for dark sectors via invisible decays of pseudoscalar $\eta$ and $\eta'$ mesons
  using the  $\pi^-$ charge-exchange reaction  on nuclei as their source have being recently reported by  NA64h, see, Ref.\cite{na64h}. 
Finally, the presented above analysis gives an illustrative order of magnitude for the sensitivity of the 
proposed experiment and may be strengthened  by more detailed 
simulations and optimisation of the beam energy and  experimental setup.
\begin{center}
{\large \bf Acknowledgments}
\end{center}
We  would like to  thank our NA64 colleagues A. Celentano, P. Crivelli, S. Kuleshov, V. Matveev, L. Molina Bueno, D.~Peshekhonov, V. Polyakov, V. Samoylenko, and A. Zhevlakov  for useful discussions, and  M. Kirsanov  for help in simulations.


\begin{thebibliography}{99}
\bibitem{referee}
We are indebted to an anonymous referee for the suggestion to estimate
   bound on the mixing parameter from the cosmological considerations.
\end{thebibliography}
\end{document}